# Shapes and energies of giant icosahedral fullerenes

## Onset of ridge sharpening transition

A. Šiber[1]

Institute of Physics, Bijenička cesta 46, 10001 Zagreb, Croatia

**Abstract.** Shapes and energies of icosahedral fullerenes are studied on an atomically detailed level. The numerical results based on the effective binary carbon-carbon potential are related to the theory of elasticity of crystalline membranes with disclinations. Depending on fullerene size, three regimes are clearly identified, each of them characterized by different geometrical properties of the fullerene shape. For extremely large fullerenes (more than about 500000 atoms), transition of fullerene shapes to their asymptotic limit is detected, in agreement with previous predictions based on generic elastic description of icosahedral shells. Quantum effects related to delocalized electrons on the fullerene surface are discussed and a simple model introduced to study such effects suggests that the transition survives even in more general circumstances.

## 1 Introduction

Ever since discovered [1], fullerene molecules continually generate interest in different scientific disciplines. This, for example, includes the predicted usage of fullerenes in (quantum [2]) computing machines [3], targeted drug delivery [4], lubricants [5] and fullerene-based chemistry [6]. One of the reasons for the interest in fullerenes is certainly their appealing geometrical form. The best known fullerene is the so-called buckminsterfullerene that contains sixty carbon (C) atoms ($C_{60}$). It is composed of twelve pentagonal carbon rings situated around the vertices of an icosahedron and twenty hexagonal carbon rings at the centers of icosahedral faces. Larger fullerenes that have an icosahedral symmetry can be constructed - these are sometimes called "giant" fullerenes [7–9]. A simple way to describe the symmetry of these molecules is illustrated in Fig. 1. The giant fullerenes can be thought of as cut-out pieces of graphene plane that are folded into a final shape (icosahedron). This procedure generates exactly twelve pentagonal carbon rings (or *disclinations*) situated around vertices of an icosahedron - all other carbon rings are hexagonal.

Soon after the prediction of giant icosahedral fullerenes, it has been noted [7] that larger fullerenes tend to deviate more from a spherical shape - the buckminsterfullerene ($C_{60}$) is nearly perfectly spherical, i.e. its atoms lie on almost equal distances from the center of the molecule. The reason for asphericity of larger icosahedral fullerenes can be quickly grasped from their construction sketched in Fig. 1. Consider a single pentagonal disclination cre-ated by cutting out a sixty degrees wedge from a circular piece of graphene sheet (see panel (c) of Fig. 1) and folding the remaining piece of the circle to rejoin the edges of a cut. It is easy to see that the shape created in such a way will be a *cone*, i.e. a pentagonal disclination shall *buckle out* from the plane of the original graphene circle. More precisely, the shape that the piece of graphene plane adopts is close to a *conical frustum* (or truncated cone) whose smaller base consists of a pentagonal ring of carbon atoms and whose half-angle is $\sin^{-1}(5/6)$. In general, the occurrence of the buckling of the sheet depends on the elastic properties of the sheet material (the bending rigidity, $\kappa$ and the two-dimensional Young's modulus or stretching rigidity, $Y$), but also on the radius (size) of the sheet ($R_s$). A shape that the elastic sheet adopts depends on the interplay between the stretching and bending contributions to the energy, i.e. a relation between $Y$, $\kappa$ and $R_s$. A detailed study of this effect has been performed in Ref. [10] and it was found that the minimal energy shape can be *uniquely* described as a function of the combined elastic parameter called the Föppl-von Kármán number (FvK), given as $\gamma = YR_s^2/\kappa$. The truncated cone is the exact solution for the shape only in the inextensional limit ($Y/\kappa \to \infty$). Tersoff has shown [11] that the energetics of icosahedral fullerenes containing less than about 2200 carbon atoms can be excellently described by accounting for the bending energy of twelve conical pieces of the fullerene surface surrounding each of the pentagonal carbon rings. This agreement strongly suggests that the surface of icosahedral fullerenes, in the regime of fullerene sizes studied by Tersoff, can be to an excellent approximation described as a union of twelve conical frusta fastened together at their larger bases (see panel (d) of Fig. 1). On the other hand, Witten and Li [12] have, on the basis of quite general scaling arguments, predicted that the *asymptotic* shape of the fullerene molecule should be a perfect icosahedron. More precisely, for very large but finite mean radius of the fullerene ($\overline{R}$), the curvature of the fullerene surface should be restricted to the region very close to the edge of an icosahedron which means that the icosahedral faces are nearly perfectly flat. The energy is localized in the icosahedral edges (ridges) along which the bending and stretching energies are of the same order of magnitude [16]. This conclusion roots in the observation that the finite curvature distributed along a large portion of the icosahedral surface (as it is the case for the union of cones) necessarily requires stretching of the sheet which becomes prohibitively expensive energywise for large fullerenes [12]. For large sizes, the minimal-energy shape of the fullerene should be thus closer to perfect icosahedron than to a geometric union of twelve conical frusta. The energy of the asymptotic shape predicted by Witten and Li also depends on the mean radius of the fullerene in a functionally different manner. Whereas the assumption of conical fullerene surface leads to logarithmic dependence of elastic energy on the mean radius ($\ln(\overline{R})$), the elastic energy of the asymptotic shape predicted by Witten and Li should scale with the mean radius as $\overline{R}^{1/3}$ (or with $N^{1/6}$, where $N$ is the total number of carbon atoms in the fullerene). This means that for some critical size of the icosahedral fullerenes, a transition in minimal energy shape of fullerene molecule should be observed. This transition has been studied for generic icosadeltahedral elastic surfaces [13,14] whose geometry can be considered as dual to icosahedral fullerene molecules [13]. The transition that effectively represents the onset of the ridge (edge) sharpening regime [15] was found for very large values of the FvK number, $\gamma = Y\overline{R}^2/\kappa > 10^6$, which means that the asymptotic shape starts to emerge for very large values of the mean radius, presuming the elastic properties of the material are fixed. Alternatively, the same effect can be observed in icosahedral shells of smaller radius whose ratio of elastic parameters, $Y/\kappa$ is very large [13].

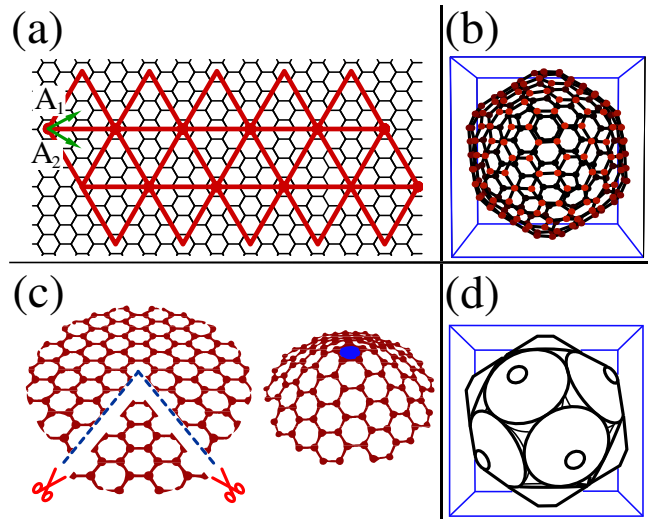

**Fig. 1.** Panel (a): A shape cut from the graphene plane - a planar construction used for generation of icosahedral fullerenes. Vectors $\mathbf{A}_1$ and $\mathbf{A}_2$ characterizing the construction are denoted. Panel (b): A relaxed fullerene shape obtained from the planar piece of graphene plane depicted in panel (a). Panel (c): An illustration of the sixty degrees wedge cut in the graphene plane and a conical shape that results once the edges of the cut are rejoined. Panel (d): An approximation of the fullerene shape shown in panel (b) by a union of twelve conical frusta whose half-angle is $\sin^{-1}(5/6)$.

## 2 All-atom numerical simulations of fullerenes: Energies

Although the ridge sharpening transition has been confirmed and thoroughly investigated in generic elastic shells [13–15], the analogous studies appropriate for fullerene molecules have not been conducted. One of the previous studies [13] indeed aimed at detecting the transition in the fullerene molecules, but the study performed was again based on generic elastic description with elasticity moduli corresponding to graphene and geometry that is dual to fullerenes (icosadeltahedron). Additional drawback of such generic studies is that the Poisson ratio is necessarily fixed to 1/3 [10], which is almost twice larger from the value pertaining to graphene [17].

An asymptotic all-atom study of fullerene molecules with the realistic interatomic potential that correctly accounts for anisotropy and nonlinearity of carbon-carbon bonding was apparently not performed and the predictions of Witten and Li have never been decisively confirmed. The reason for the lack of such study lies in the fact that the transition towards the asymptotic shape is observed for fullerenes containing extremely large number of carbon atoms, as shall be shown in this article. Thus, the atomically detailed calculations become very difficult and slow.

The interactions of carbon atoms in fullerene molecules are modeled in this article using the second-generation reactive empirical bond order potential energy expression recently proposed by Brenner and coworkers [19]. This relatively simple potential accounts for anisotropy of carbon-carbon interactions, as well as many-body effects. Note that regardless of its relative simplicity, the potential still contains many features that are not present in the linear and nearest-neighbor-only model of interaction adopted for generic icosadeltahedral elastic shells [13,14], and it predicts the realistic values for elastic moduli of graphene (see below; the Poisson ratio is predicted to be 0.19).

The icosahedral fullerenes are constructed as illustrated in panel (a) of Fig. 1. Briefly, the outlined shape is cut from the graphene plane and folded into an icosahedron and the edges are 'glued' together (i.e. molecular bonds are reestablished between the nearest neighboring carbon atoms). The thus obtained perfect icosahedral shape is used as a starting guess for the minimal-energy shape. The shape is relaxed using a particularly efficient implementation of the conjugate gradient algorithm [18] until a solution (shape) that minimizes the total potential energy is found. The minimal energy shape that is obtained using a planar construction shown in panel (a) of Fig. 1 is displayed in panel (b) of Fig. 1. Note that in this case, the shape can be excellently approximated by a union of twelve conical frusta whose apices are located at the position of icosahedron's vertices (panel (d) of Fig. 1). As discussed earlier, the conical shape results from the tendency of the graphene plane to buckle out in vicinity of pentagonal disclinations, as shown in panel (c) of Fig. 1. The size of the icosahedral shape constructed can be varied depending on the lengths of vectors $\mathbf{A}_1$ and $\mathbf{A}_2$ shown in panel (a) of Fig. 1. These are integer multiples of $a\sqrt{3}$ where $a$ is the nearest neighbor carbon-carbon separation ($a = 1.4204$ Å for the Brenner potential [19]), i.e. $|\mathbf{A}_1| = |\mathbf{A}_2| = ma\sqrt{3}$, $m = 1, 2, 3, \ldots$. The total number of atoms ($N$) in the thus constructed shapes is $N = 60m^2$ [9].

Panel (a) of Fig. 2 displays the excess energies ($\Delta E$) of giant icosahedral fullerenes as a function of the total number of atoms that they contain. These are obtained as the difference between the calculated energies of minimal energy shapes and the energies that the same number of carbon atoms would have in the infinite graphene plane. The excess energies originate from combined effects of 'core' disclination energies, which are local energies associated with the pentagonal carbon rings, and the elastic energies related to the curvature and stretching that is present in the fullerene surface [11,9]. The Brenner potential predicts that the energy per atom in the infinite graphene plane is -7.3949362728 eV. The high numerical precision of this energy is necessary since the effects that are sought in this article turn out to be rather subtle. Note that for fullerenes that contain up to about $10^5$ atoms, the excess energy perfectly scales with the logarithm of total number of atoms ($N$) in accordance with Tersoff's prediction and this indeed corroborates a simple geometrical picture of a fullerene molecule as a union of twelve conical frusta. This also means that practically all of the elastic energy is of the bending type. For larger fullerenes a clear deviation from the predicted logarithmic scaling is observed (see inset in Fig. 2a). Note, however, that the error one would make in estimating the excess energy from $\Delta E \sim \ln(N)$ would be about 5 eV for the largest fullerene studied ($N = 1261500$). The total energy of this fullerene is $-9.32862 \times 10^6$ eV, and the error is thus completely negligible on the scale of total energy. The effect sought for is thus rather subtle, as announced, and high precisions are needed to detect it.

## 3 All-atom numerical simulations of fullerenes: Shapes

Although the study of excess energy clearly demonstrates that a kind of a slow transition takes place for large fullerenes ($N > 10^5$), more insight can be obtained by studying geometrical properties of the minimal energy shapes. I have chosen the mean square asphericity ($\sigma$) as a *global* measure of the shape geometry. This quantity is calculated as

$$\sigma = \frac{\langle \Delta R^2 \rangle}{\langle R \rangle^2} = \frac{1}{N} \sum_{i=1}^{N} \frac{(|\mathbf{r}_i - \mathbf{r}_0| - \langle R \rangle)^2}{\langle R \rangle^2}, \qquad (1)$$

where $\mathbf{r}_0$ is the vector of the geometrical center of the shape, $\mathbf{r}_i$ is the vector of $i$-th carbon atom in the shape, $i = 1, \ldots, N$, and $\langle R \rangle$ is the mean radius of the shape, $\langle R \rangle = \sum_{i=1}^{N} |\mathbf{r}_i - \mathbf{r}_0|/N$. The same measure of asphericity was used in Refs. [14,20] which enables a reliable comparison of the present results with those pertaining to generic icosadeltahedral elastic shells. Panel (b) displays the mean square aspherities of fullerenes as a function of FvK number ($\gamma$) which was calculated as $Y\langle R\rangle^2/\kappa$, where stretching ($Y$) and bending ($\kappa$) moduli correspond to infinite graphene plane and their values can be easily calculated from the Brenner's potential which yields $Y = 23.35$ eV/Å$^2$ and $\kappa = 0.83$ eV [9]. Note how $\sigma$ increases from practically zero for buckminsterfullerene ($m = 1$) and saturates at about $\sigma \approx 0.016$ for $2000 < N < 120000$. Then another increase in asphericity is observed for $N > 120000$ i.e. $\gamma > 10^6$. Characterization of fullerene shapes with their corresponding FvK number, strictly speaking, makes sense only for very large fullerenes for which a continuum description is applicable [9]. Nevertheless, Fig. 2 clearly shows a good agreement of the present results (squares) with those obtained for generic elastic shells in Ref. [20] (full line) which are essentially the same as in Ref. [14]. In those calculations, the number of vertices in the shell was kept constant and the change in FvK number originates from variation of bending rigidity $\kappa$, whereas in case of fullerenes, an increase of the fullerene radius (number of atoms) generates the change in $\gamma$. The observed agreement further corroborates the conclusion that the detected nonlinear dependence of excess energy on $\ln(N)$ signify the *onset of ridge sharpening transition* that is found in generic elastic shells for $\gamma > 10^6$, i.e. in the same interval as found here for the case of fullerenes. Additional indication of the ridge sharpening regime can be found from the functional behavior of excess energy on the total number of atoms. The inset in Fig. 2a displays the fit of excess energies to $C_1 N^{1/6} + C_2$ behavior which yields $C_1 = 5.595$ eV and $C_2 = 33.38$ eV. It is interesting to note that the factor $C_1$ is only about 33 % larger from the value that would be obtained from the sum of the energies of 30 icosahedral ridges in the asymptotic regime as predicted numerically in Refs. [15,16]. The energy of a single ridge was found to be given by [15,16]

$$E_{ridge} = 1.24 \left(\frac{YL^2}{\kappa}\right)^{1/6} \alpha^{7/3}, \qquad (2)$$

where $L$ is the length of the ridge, and $\pi - 2\alpha$ is the dihedral angle of the ridge. Inserting values appropriate for the icosahedral carbon shape ($\alpha = 0.365$ rad - the angle characteristic of an icosahedron and elastic parameters $Y$

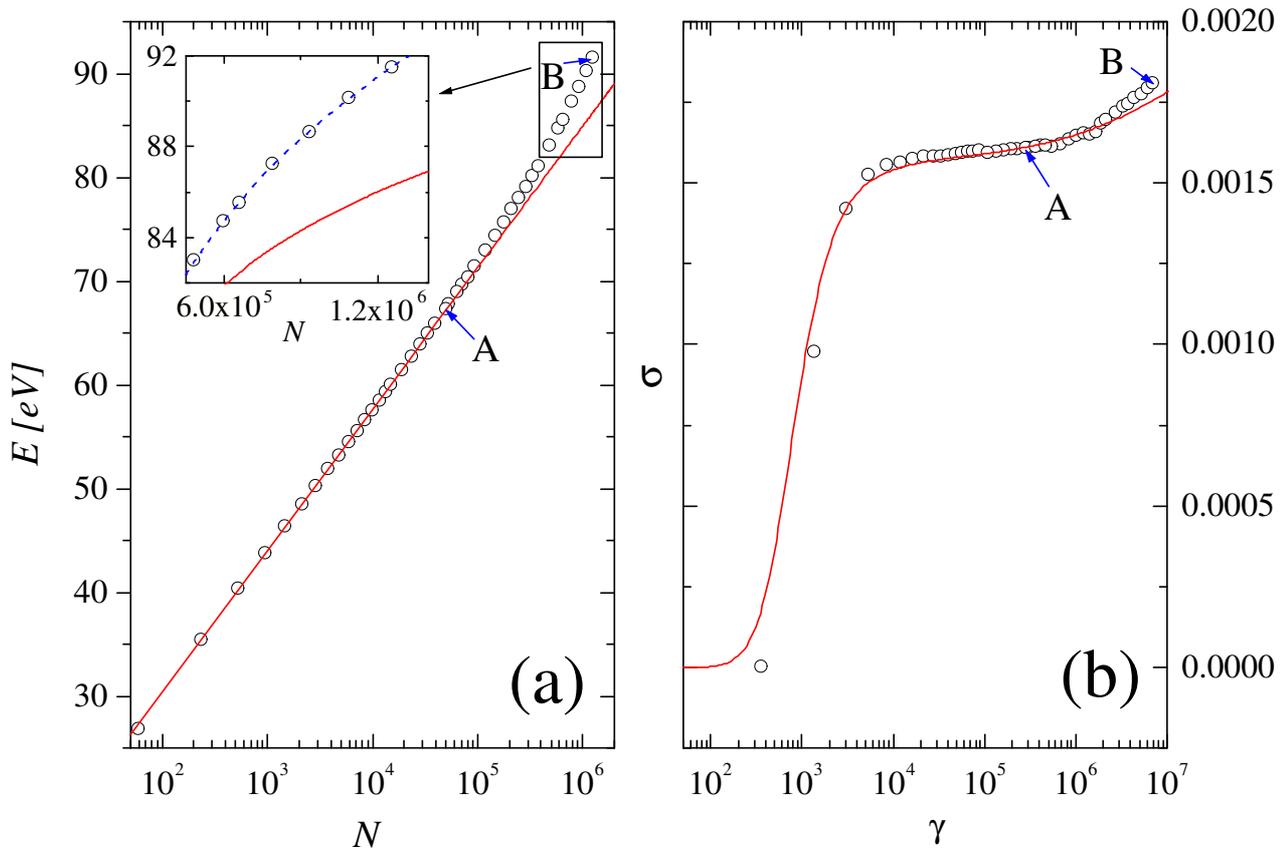

**Fig. 2.** Panel (a): Excess energy of icosahedral fullerenes as a function of total number of carbon atoms. Empty circles are results of numerical minimization, while the full line shows the prediction of the theory of elasticity based on the assumption that the shapes can be described as unions of twelve conical frusta. The inset shows a more detailed view of the excess energy in the region where $N > 450000$. The dashed line in the inset displays $N^{1/6}$ dependence of the excess energy (see text). Panel (b): Mean square asphericity as a function of Föppl-von Karman number characteristic of fullerene structures. The full line shows the results for generic icosadeltahedral shells studied in Ref. [20].

and $\kappa$ as specified above) and summing over all 30 ridges of an icosahedron, one finds that

$$E_{icosahedron} = 4.201[\text{eV}] N^{1/6}. \qquad (3)$$

The difference between the two values (5.595 eV obtained numerically here vs. 4.201 eV extracted from numerical studies of a single elastic ridge in Ref. [15]) may be related to the fact that 5 ridges end in the same vertex (unlike in Ref. [16] where a single ridge bounded by two vertices is studied). Additionally, the Poisson ratio of graphene differs significantly from 1/3 which is also expected to somewhat influence the multiplicative constant of the scaling relation for the ridge energy [21]. Constant $C_2$ is close to, albeit somewhat larger from the sum of twelve core energies of pentagonal disclinations [9].

In addition to asphericity, which is a global measure of the shape geometry, it is of use to examine local geometrical properties of the shape. For further discussion, it is of use to fix the coordinate system for description of the shape. For this purpose, the shape is rotated in such a way that the z-axis of the Cartesian coordinate system coincides with one of shape's six C5 axes of symmetry. A perfect icosahedron can be constructed so that its twelve vertices coincide with the centers of the shape's carbon pentagonal rings. Two vertices of this icosahedron lie on the z-axis and the one with larger z-coordinate is denoted as T (see Fig. 3). Vertex T is an apex of the pentagonal pyramid whose basis is defined by the five icosahedron vertices that are nearest neighbors of T. The pentagonal pyramid is cut by three planes denoted by $\pi_1$, $\pi_2$, and $\pi_3$ in panel (c) of Fig. 3, which are chosen so that their distance from the topmost vertex (T) is one third, ($\pi_3$) one half ($\pi_2$), and a whole pyramid height ($\pi_1$). All the carbon atoms whose distance from a particular plane is less than $a/2$ are then projected onto a plane, which yields a two-dimensional cross-section of the shape. There are thus three cross-sections that are examined. Panels (a) and (b) of Fig. 3 display the three cross-sections for the shapes denoted by A and B in Fig. 2, respectively. Note the completely circular $\pi_3$ cross-section of shape A. This clearly demonstrates that the region around the disclination in shape A is indeed a cone. Unlike the shape A, shape B is in the ridge sharpening regime (see Fig. 2) which can be easily seen from pentagonal nature of *all* of the cross-sections. This means that the curvature of the fullerene surface accumulates around the ridges, i.e. the icosahedral faces are nearly flat, as expected in the ridge sharpening regime [12]. Note that the linear dimensions of the cross-sections of shape B are about five times larger from the corresponding cross-sections of shape A. A (small) deviation from the circular shape of the $\pi_2$ cross-section is to be expected even for shape A, since the $\pi_2$ cross-section goes exactly around the lowest base of a cone (whose apex is close to point T), exactly in the region where the 6 cones should be "fastened together". The $\pi_1$ cross-section goes along the shape ridges, and if the shape is indeed properly describable by union of cones, the cross-section should be close to a perfect pentagon. In fact, an inward stretching of the ridges is easily observed, particularly for shape A which contributes to the stretching energy of the shape. It is precisely this effect that induces the ridge sharpening transition when the shape becomes large [12].

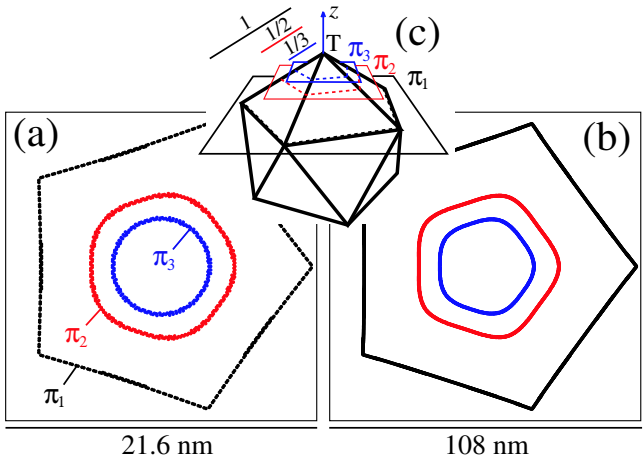

**Fig. 3.** Panel (a): Cross-sections of $m = 29$ shape ($N = 50460$). Panel (b): Cross-sections of $m = 145$ shape ($N = 1261500$). Panel (c): Illustration of the procedure used to display cross-sections in panels (a) and (b).

## 4 Limitations of the study: electron delocalization effects

Although the numerical results clearly demonstrate the onset of ridge-sharpening transition, one has to keep in mind that this was proven only for the *model* fullerenes, i.e. those in which the carbon atoms interact as prescribed by the bond-order potential [19]. Although this model includes the many-body effects in C-C interaction, effectively only the short-range interactions between the carbon atoms (second-nearest-neighbor) are accounted for. One may wonder whether the extremely subtle nature of the ridge-sharpening transition will be influenced by the delocalization of electrons over the *whole* fullerene molecule and by the corresponding consequences on the total energy of the molecule. It seems rather difficult to precisely account for this effect in carbon sp$^2$ structures even with the present state-of-the-art total electronic energy calculations. However, a simple argument suggests that the ridge-sharpening transition may survive even when one includes the effects of the electron delocalization. It seems reasonable to approximate the fullerene surface as a *geometrical confinement* acting on the delocalized electrons in it, i.e. the dynamics of the delocalized electrons is *constrained* to evolve on the fullerene surface while the "residual" dynamics related to degrees of freedom perpendicular to the surface is frozen. One is thus interested in the effects that the details of the fullerene shape may have on the energy of the electrons constrained to "live" on the fullerene surface. Quantum mechanics of particles in constrained geometries has been studied previously [22–24] and it was found that the three-dimensional Schrödinger equation for the constrained particles reduces to Schrödinger-like equation in which the mean and gaussian curvatures of the constraint (surface) appear as effective potentials. The geometry of the fullerene is fairly complex, nevertheless a clear insight may be obtained by concentrating on a *single* fullerene ridge. It helps to isolate a particular ridge and approximate it as a (open) "book-cover" shaped surface obtained by bending a plane around the cylinder of radius $R$ [22] (see Fig. 4). It has been shown [22] that the stationary Schrödinger equation for a single particle on such a surface reduces to

$$-\frac{\hbar^2}{2M}\left(\frac{\partial^2 \xi(z,s)}{\partial s^2} + \frac{\partial^2 \xi(z,s)}{\partial z^2}\right) - \frac{\hbar^2 K(s)^2}{8M}\xi = E_{(z,s)}\xi(z,s), \quad (4)$$

where $\xi$ is the wavefunction of the particle on the surface, $s$ is the arc-length coordinate of the cross-section of the surface (perpendicular to the ridge), $z$ is the cartesian coordinate parallel to the ridge direction (see Fig. 4), $M$ is the particle mass, and $E_{(z,s)}$ is the particle energy related to its dynamics on the constraint that is defined by $z$ and $s$ coordinates. The mean curvature of the bookcover surface depends only on $s$ coordinate and is denoted by $K(s)$ in the above equation. As mentioned before, the squared curvature appears as effective potential, $V(s) = -\hbar^2 K(s)^2/8M$. The total energy ($E$) of the particle on the constraint is easily found to be

$$E = \frac{\hbar^2 k_z^2}{2M} + E_s + E_{perp}, \quad (5)$$

where $k_z$ is the wave-vector of the particle propagating in the $z$-direction, $E_s$ is the eigenvalue of the effective (separated) one-dimensional Schrödinger equation in $s$-coordinate, and $E_{perp}$ is the "confinement" energy, i.e. an energy required to confine the quantum particle in the directions perpendicular to the surface [22]. A one-dimensional eigenvalue $E_s$ can be found as the solution of the Schrödinger equation for a particle bound by a finite square well potential ($K(s) = 1/R$ in the cylindrical portion of the bookcover surface and is zero otherwise)

$$V(s) = \begin{cases} 0 & , |s| > R\alpha \\ V(R) = -\hbar^2/(8MR^2) & , |s| < R\alpha \end{cases} \quad (6)$$

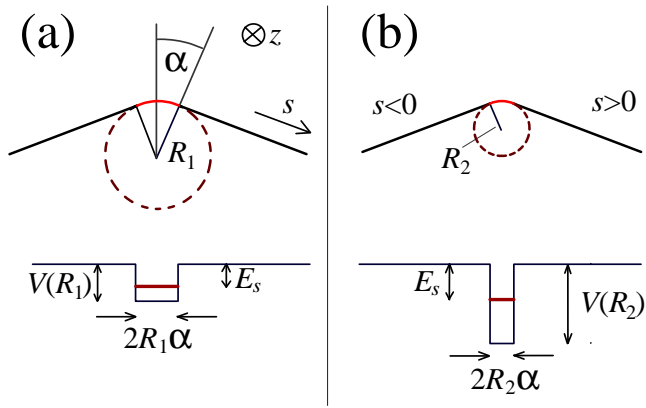

**Fig. 4.** A model of the ridge obtained by bending a plane around the surface of a cylinder. Panels (a) and (b) represent the ridges with different curvatures with the same dihedral angle and the effective one-dimensional potentials discussed in the text. Parameters, physical quantities and coordinates relevant to the discussion in the text are also denoted.

The solution of the problem is described in many textbooks on quantum mechanics (see e.g. Ref. [25]) and one finds that for $\alpha < \pi$ (as is in the case of interest) the potential supports only one bound state (localized in the ridge region) whose energy is given by

$$E_s = -\frac{\hbar^2}{2MR^2\alpha^2}\eta^2, \quad (7)$$

where $\eta$ is the solution of the system of equations

$$\xi \tan \xi = \eta$$
$$\xi^2 + \eta^2 = \alpha^2/4. \quad (8)$$

The angle ($\alpha$, see Fig. 4) of the bookcover surface is fixed to $\alpha = 0.365$ rad, so that $\pi - 2\alpha$ is the dihedral angle at the icosahedron ridge. In this case, $\eta = 0.0326$ (irrespectively of $R$) and the bound state energy is thus $E_s = -0.00106\,\hbar^2/(2MR^2\alpha^2)$. Consider now two bookcover surfaces with the same opening angle $\alpha$ whose ridges differ in curvature, as sketched in panels (a) and (b) of Fig. 4. Since the binding energy of the electron for fixed $\alpha$ scales as $R^{-2}$ [26] (see Eq.(7)), a sharper ridge (i.e. the one with smaller $R$, panel (b) of Fig. 4) binds the electron more strongly (note that the other two contributions to the binding energy, $E_z$ and $E_{perp}$ are the same, irrespectively of the ridge sharpness). Thus, this simple model suggests that the binding energy of the free electron gas *increases* in the process of ridge sharpening transition, i.e. the transition is favored energetically by the free delocalized charge residing on the ridge surface. Of course, the model introduced cannot be directly transcribed to the case of fullerenes, nevertheless it does suggest that the ridge-sharpening transition may survive and even be supported by the delocalized part of the electronic density.

## 5 Summary and Conclusion

I have presented all-atom calculations of the shapes and energies of giant icosahedral fullerenes. It has been shown that the atomically-resolved calculations support the findings obtained from the study of continuum elasticity theory of membranes [12] with some minor corrections, most of which can be related to the Poisson ratio of the graphene surface being significantly different from 1/3. In particular, the ridge-sharpening transition towards the asymptotic fullerene shape predicted previously in Ref. [12] was found to occur also in the atomically detailed calculations. Limitations of the calculations are discussed, in particular with respect to quantum effects resulting from the electronic delocalization on the fullerene surface that the adopted model of carbon-carbon interactions [19] cannot fully describe. Using a very simplified model of the electronic delocalization, it was found that the ridge-sharpening transition may even be favored by such effects.

## References

1. H. W. Kroto, J. R. Heath, S. C. O'Brien, R. F. Curl and R. E. Smalley R. E., Nature **318**, 162 (1985)
2. W. Harneit, Phys. Rev. A **65**, 032322 (2002)
3. D. Tománek, J. Phys.:Condens. Matter **17**, R413 (2005)
4. L. Mazzola, Nature Biotechnology **21**, 1137 (2003)
5. T. Coffey, M. AbdelMaksoud and J. Krim, J. Phys.:Condens. Matter **13**, 4991 (2001)
6. P. R. Birkett, Annu. Rep. Prog. Chem., Sect. A **95**, 431 (1999)
7. H. W. Kroto and K. McKay, Nature **331**, 328 (1988)
8. S. Itoh, P. Ordejón, D. A. Drabold and R. M. Martin, Phys. Rev. B **53**, 2133 (1996)
9. A. Šiber, Nanotechnology **17**, 3598 (2006)
10. H. S. Seung and D. R. Nelson, Phys. Rev. A **38**, 1005 (1988)
11. J. Tersoff, Phys. Rev. B **46**, 15546 (1992)
12. T. A. Witten and H. Li, Europhys. Lett. **23**, 51 (1993)
13. Z. Zhang, H. T. Davis, R. S. Maier and D. M. Kroll, Phys. Rev. B **52**, 5404 (1995)
14. J. Lidmar, L. Mirny and D. R. Nelson, Phys. Rev. E **68**, 051910 (2003)
15. A. E. Lobkovsky, Phys. Rev. E **53**, 3750 (1996)
16. A. E. Lobkovsky and T. A. Witten, Phys. Rev. E **55**, 1577 (1997)
17. B. I. Yakobson and Ph. Avouris, *Carbon nanotubes*, Topics Appl. Phys. **80**, eds. Dresselhaus M. S., Dresselhaus G., and Avouris Ph. (Springer-Verlag, Berlin Heidelberg, 2001)
18. W. W. Hager and H. Zhang, Siam J. Optim. **16**, 170 (2005)
19. D. W. Brenner, O. A. Shenderova, J. A. Harrison, S. J. Stuart, B. Ni and S. Sinnott, J. Phys.: Condens. Matter **14**, 783 (2002)
20. A. Šiber, Phys. Rev. E **73**, 061915 (2006)
21. A. E. Lobkovsky, private communication.
22. R. C. T. da Costa, Phys. Rev. A **23**, 1982 (1981)
23. J. Gravesen, M. Willatzen, and L. C. Lew Yan Voon, J. Math. Phys. **46**, 012107 (2005)
24. J. Gravesen and M. Willatzen, Phys. Rev. A **72**, 032108 (2005)
25. L. I. Schiff, *Quantum mechanics*, 3rd edn. (McGraw-Hill, 1968)
26. A. V. Chaplik and R. H. Blick, New J. Phys. **6**, 33 (2004)